\documentclass[preprint,12pt]{elsarticle}

\usepackage{amssymb}
\usepackage{amsmath}
\usepackage{amsthm}
\usepackage{amsfonts}
\usepackage{amssymb}
\usepackage{xcolor,graphicx}
\usepackage{tikz}
\usepackage{color}
\usepackage{longtable}
\usepackage{array}
\usepackage{dsfont}

\newcounter{bla}

\journal{Computer Physics Communications}

\begin{document}

\begin{frontmatter}



\title{myqcs: Quantum Circuit Simulator}

\author[a]{J. P. J. Rodriguez\corref{author}}

\cortext[author] {Corresponding author.\\\textit{E-mail address:} drjpjrodriguez@gmail.com}
\address[a]{School of Mathematics, Statistics \& Applied Mathematics, NUI Galway, Ireland}

\begin{abstract}
myqcs is a quantum circuit simulator using tensor. The simulator can be used with both wave functions or density matrices. The simulator also included noise that can be varied in form and that can be different for each gate. 

\end{abstract}

\begin{keyword}
Quantum Circuit; Quantum Noise; NISQ; Simulator;

\end{keyword}

\end{frontmatter}



{\bf PROGRAM SUMMARY}

\begin{small}
\noindent
{\em Program Title: myqcs}                                          \\
{\em CPC Library link to program files:} (to be added by Technical Editor) \\
{\em Developer's repository link:github.com/smeeton1/myqcs} \\
{\em Code Ocean capsule:} (to be added by Technical Editor)\\
{\em Licensing provisions: GPLv3}  \\
{\em Programming language: Julia}                                  \\
{\em Nature of problem: Simulation of quantum circuits in both a density matrix and wave function form.}\\
{\em Solution method: A tensor method is used to solve quantum circuits in both the density matrix and wave function form.}\\
   \\

\section{Introduction}

Currently we are entering the start of the quantum computing era as various quantum computers and accelerators begin to enter the market. These machines are often referred to as  noisy intermediate-scale quantum machines\cite{Preskill2018quantumcomputingin} (NISQ). Some of these accelerator have begun to have hundreds of qubits, though this still remaining below the amount needed for error correction techniques\cite{RevModPhys.94.015004}.  These devices are limited in performance by the noise on the qubits causing decoherence\cite{Leymann_2020}. Due to these limitations the circuits run on these devices are often required to be small and have limited depth. 

The Limitations of NISQs mean that efficient algorithms are needed to get the most out of them. This is one of the places that quantum circuit simulator has become a vital tool. As they provide a way to test such code with out the need to use the few resources available or for the long waits in queues\cite{9137029}. The simulators can also be used as a test ground for software meant to run quantum accelerators\cite{10.1145/3474222}.

Simulator have also found a place in the parallelization of quantum algorithms as well. A tensor simulation can be used to determine the best slicing point in the circuit. The simulator can be used to compute the SVD of the circuit and then determine the qubits with the lowest bond rank\cite{9537173}. The simulator also provide a place to test sub-circuit setups\cite{10.1145/3466752.3480116}. Simulators also have a role in helping to place small circuits efficiently on to larger accelerator\cite{10.5555/3539845.3540090,PhysRevLett.120.210501}.
 
The noise found in these machines has a large effect on the out come of the circuits being run. This creates the need to simulate and characterize this noise. There have been various methods for with some characterize the noise found on particular machines\cite{PhysRevA.103.042603}. Many other have used general models for the noise\cite{Noh2020efficientclassical,SNQC,Chen2021,Fujii2016}. 

As the number of available quantum accelerator have been increasing along with access to then, so has the number of simulator for them increased. A list of some simulator can be found in the ref.\cite{OSQSP}. Many of these simulators have use tensors as tensor networks have been found to be useful tools in solving quantum systems\cite{10.5555/3433701.3433719,Wood2015TensorNA}. The most popular of these simulators being Qiskit\cite{gadi_aleksandrowicz_2019_2562111}. 

Here we describe a tensor based quantum circuit simulator with noise. The simulator has been programmed to work in a jupyter notebook as well as having a sand alone command line application. This simulator has been given a wide range of options for the user and can work with the most common of quantum scripting language OpenQASM\cite{openqasm}. 

\section{Theoretical background}

The gate in a quantum circuits can be treated as a tensor \cite{PhysRevA.102.062614,doi:10.1137/050644756,10.1371/journal.pone.0206704,PhysRevLett.119.040502} and the circuit can be written as a network of these tensor. Here we use this treatment to simulate the circuit. In this method the state of the qubits is treated as a 1D tensor and the gate are represented as 2D tensors. Once the circuit is written in this form various method can be used to contract the tensor network to get the output of the qubits.

Three methods are included in the package. There is a simple method that combines all the qubits and then applies the gates as they where added to the circuit. 

The second method is an matrix product state(MPS) \cite{PhysRevA.101.032310,mpor,ORUS2014117}. In this method the the circuit is split according to the qubits. Each qubit is then solved separately and finally they are combined in the last step.

The last method is a depth controlled method. Here the gate are applied according to the depth of the circuit. The qubit are combined as multi-qubit gates are applied. Any uncombined qubits are combined in the last stage. This method can also be set to stop at any depth to see the results of part of the circuit.  

\subsection{Measurement}

Included in the package is the ability to measure qubits\cite{PhysRevA.64.052312,wiseman_milburn_2009}. To get the results of a measurement, first the qubit of interest is separated from the mixed state using a partial trace. The measurement is then performed and the resulting state is then recombined with the total state again.   

The measurements can be placed any where in the circuit and the results are written to a classical bit chosen by the user. The classical bits can then be used to control other gates. 

\subsection{Noise}

Noise has been integrated into the package and can be added in two ways. The noise can be added to all gate or just a selection of the gates. Included in the package is dephasing, depolarizing, and amplitude damping noise. For the definitions of the noise $\varepsilon$ acting on the state $\rho$ we followed the treatment of Cheng et.al. \cite{SNQC}. We start with the Kraus representation 
\begin{equation}
\label{kraus}
\varepsilon (\rho) = \Sigma_k U E_{k} \rho E_{k}^{\dagger}  U^{\dagger},
\end{equation}
where the $E_{k}$'s are the Kraus operators and we have the requirement $\Sigma_k E_{k} E_{k}^{\dagger} = I$. We can then write the noise for one gate as 
 
\begin{equation}
\label{noise}
 \rho \rightarrow U \circ \varepsilon (\rho).
\end{equation}

We begin by define the dephasing noise as\cite{SNQC} 
\begin{equation}
\label{dephasing}
\varepsilon_{Ph}(\rho) = (1-\epsilon)\rho + \epsilon Z \rho Z^{\dagger},
\end{equation}
with Z being the Pauli operator and $\epsilon \in [0,1] $. We will then define the depolarizing noise as 
\begin{equation}
\label{depolarizing}
\varepsilon_{Po}(\rho) = (1-\epsilon)U\rho U^{\dagger}+ \epsilon \frac{I}{2}.
\end{equation}
Finally we will define the amplitude damping as
\begin{equation}
\label{amplitude}
\varepsilon_{A}(\rho) = A_{0}\rho A_{0}^{\dagger}+ A_{1}\rho A_{1}^{\dagger}
\end{equation}
with $A_{0}= \begin{pmatrix} 
                1 & 0 \\
                0 & \sqrt{1-\epsilon}
             \end{pmatrix}$
and $A_{1}= \begin{pmatrix} 
                0 & \sqrt{\epsilon} \\
                0 & 0
             \end{pmatrix}$.

In the case of a two bit gate the noise will be the tensor product of the noise for each qubit as shown below 
\begin{equation}
\label{nosie2bit}
U \circ \varepsilon_{Q1} \otimes \varepsilon_{Q2} (\rho).
\end{equation}

\section{Computational details}

All the method included in myqcs can be used with both the wave function or the density matrix. The circuit can be input directly in the jupyter notebook or by importing an OpenQASM\cite{openqasm} file.

For bench-marking and testing there is a random circuit function. The circuit generated can be set to any number of qubits and depth. the circuit alternates between single qubit gates and two qubit gates. The random single qubit gates are applied to each qubit and the two qubit gates are applied between closest neighbors. The two qubit gates are shifted so that all closest neighbors are connected. An example of a generic circuit can be found in Fig. \ref{random}.

\begin{figure}
 \centering
 \includegraphics[scale=0.5]{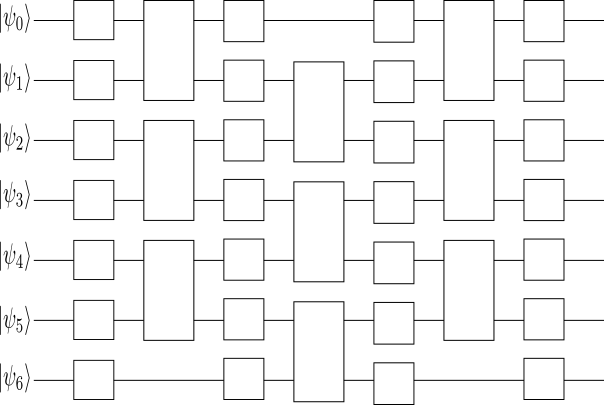}
 \caption{Example of the pattern used to generate a random circuit.}
 \label{random}
\end{figure}

The random circuit function above was used to benchmark the solving methods included in the package. The circuits where set to 10 qubits and the depth was varied from 5 to 30. The results for the bench-marking are shown in Fig. \ref{t_graph}. The results for the simple method are shown in Fig. \ref{t_graph} (a), the MPS method is shown in Fig. \ref{t_graph} (b), and finally the depth method is shown in Fig. \ref{t_graph} (c). The best time were gotten using the MPS method which had very little variation over the range shown. The depth performs the worst, which is caused by the requirements to perform the gates in order of depth. This method like the simple method show a growth in the time needed to solve the circuit that corresponded with the depth of the circuit. The simple method show a liner growth while the depth method show exponential growth.

\begin{figure}
 \centering
 \includegraphics[scale=0.5]{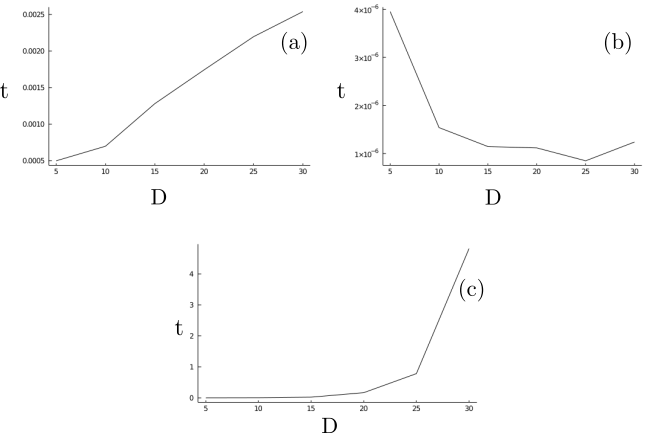}
 \caption{The time(t) need for the (a) simple solver, (b) MPO methods, and by (c) the depth method, as a function of depth(D).}
 \label{t_graph}
\end{figure}

To evaluate the noise the fidelity between mixed states was used \cite{Liang_2019}. The fidelity for two mixed states is defined as 
\begin{equation}
 F=(tr\sqrt{\sqrt{B}A\sqrt{B}})^2.
\end{equation}
With $A$ and $B$ being the two density matrices being compared. For noise to be applied to the circuit, it is required that the simulation be run using density matrices.

The results for the fidelity with the noise included in the package are shown in Fig. \ref{f_graph}. To generate the curves a random circuit of 5 qubits with a depth of 15 were used. $\varepsilon$ was varied between 0.1 and 0.9. The fidelity for the depolarizing noise is shown in Fig. \ref{f_graph} (a), the  dephasing noise is shown in Fig. \ref{f_graph} (b), and the amplitude damping noise is shown in Fig. \ref{f_graph} (c). As can be seen all the forms of noise cause the fidelity to drop with the depolarizing noise being the fastest. The dephasing noise gives a plateau before dropping off and the amplitude damping noise goes down linearly.

\begin{figure}
 \centering
 \includegraphics[scale=0.5]{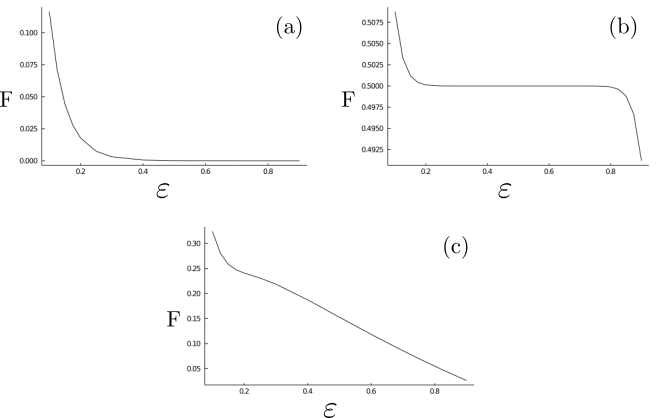}
 \caption{The Fidelity(f) as it varies with $\varepsilon$ for (a) depolarizing noise, (b) dephasing noise, and (c) amplitude damping noise.}
 \label{f_graph}
\end{figure}

\section{Applications}

myqcs provides a quantum circuit simulator with noise. The noise can be added to only the gates desired and can be varied for each gate separately or be made uniform. The main application for this package is the simulation of quantum circuits with or with out noise. The software can also be applied when unique noise is required for each gate.

\section{Conclusion}

Here we present a versatile quantum circuit simulator, myqcs. The simulator can be used with both density matrices and wave functions. The package is also capable of handling noisy gates. The package can handling measurements and measurement controlled gates as well.

\end{small}






\bibliographystyle{elsarticle-num}







\end{document}